\documentclass{article}
\usepackage{spconf,amsmath,graphicx}
\usepackage{float}
\usepackage{algorithm2e, comment}

\RestyleAlgo{ruled}

\title{Adaptive Endpointing with Deep Contextual Multi$-$armed Bandits}
\name{\parbox{\textwidth}{\centering
Do June Min\thanks{*This work was done during the author's Amazon internship.}${^1}{^*}$,  Andreas Stolcke${^2}$,  Anirudh Raju${^2}$, Colin Vaz${^2}$, \\
Di He${^2}$, Venkatesh Ravichandran${^2}$ and  Viet Anh Trinh${^2}$}}
\address{${^1}$University of Michigan \quad ${^2}$Amazon Alexa AI, USA \\
dojmin@umich.edu \quad  \{stolcke, ranirudh, vazcoli, deehe, veravic, trinhvie\}@amazon.com}

\begin{document}
\ninept
\maketitle{}
\begin{abstract}
Current endpointing (EP) solutions learn in a supervised framework, which does not allow the model to incorporate feedback and improve in an online setting. Also, it is a common practice to utilize costly grid-search to find the best configuration for an endpointing model. 
In this paper, we aim to provide a solution for adaptive endpointing by proposing an efficient method for choosing an optimal endpointing configuration given utterance-level audio features in an online setting, while avoiding hyperparameter grid-search.
Our method does not require ground truth labels, and only uses online learning from reward signals without requiring annotated labels.
Specifically, we propose a deep contextual multi-armed bandit-based approach, which combines the representational power of neural networks with the action exploration behavior of Thompson modeling algorithms.
We compare our approach to several baselines, and show that our deep bandit models also succeed in reducing early cutoff errors while maintaining low latency.
\end{abstract}
%

\section{Introduction}
\label{sec:intro}

In modern spoken language AI assistants and dialog systems, endpointing is a key step in the system pipeline, determining when a speaker has finished an utterance \cite{Ferrer2003, Arsikere2014ComputationallyefficientEF, Li2022, Chang2022}.
Similar to turn-taking in human-human conversations, smooth endpointing that avoids early cutoffs of speaker utterances or excessive latency before an agent response is key to efficient conversational interaction \cite{Schlangen2006FromRT}.
For instance, speech disfluencies in the form of pauses can lead to poor endpointing, and require attention to prosodic properties to avoid mistaking them for utterance-final pauses \cite{Arsikere2014ComputationallyefficientEF}.
Regardless of the modeling used, endpointing hyperparameters need to be carefully calibrated, e.g., to find a good balance between early cutoffs and latency \cite{Zhao2021, Lu2022, Huang2022E2ESJ}.


To simplify the problem for learning purposes, in this paper we investigate learning the choice 
between just two endpointing configurations, ``standard'' and ``relaxed'', using features that are extracted for each speaker or utterance (i.e., a sequence of utterances).
Whereas the ``standard'' configuration leads to endpointing behavior suitable for most speakers, the ``relaxed'' configuration is suited for utterances with slow speaking rate and more mid-utterance pausing.
Thus, the task of endpointing adaptation is formulated as finding the better configuration for each utterance.
Although considerable work has been done on different endpointing models and algorithms, there have been few studies on how endpointing hyperparameters can be optimized at a personal or contextual level \cite{Maas2018, Jayasimha2021PersonalizingSS, Ding2022PersonalV2}.
Furthermore, adaptive decision making in acoustic modeling has been studied mostly in the context of ASR \cite{Munkhdalai2022, Sathyendra2022}, rather than endpointing.

Specifically, our goal is to address the following questions:
\begin{figure}[t]
  \centering
  \includegraphics[width=\linewidth]{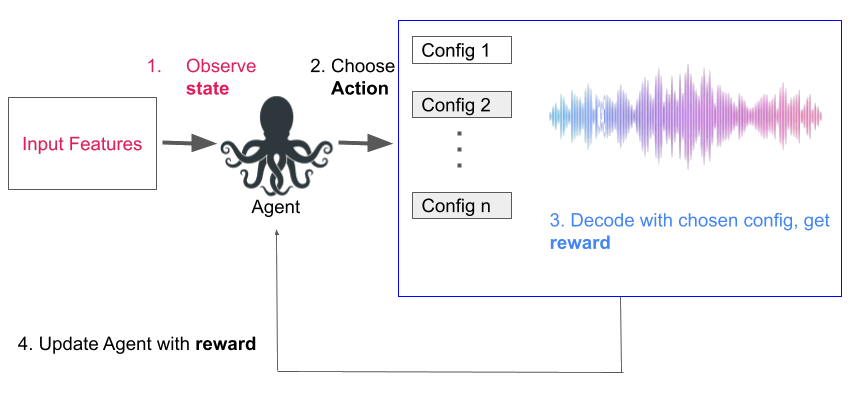}
  \caption{An overview of our adaptive endpointing with deep contextual multi-armed bandit (CMAB).}
  \label{fig:bandit_diagram}
\end{figure}

\noindent\textbf{Which features are most useful?} 
As inputs to the adaptive endpointing model, different types of features can be considered.
Features that can be extracted directly from the target audio are good candidates.
Although it is not realistic to assume the full target audio would be available in time for adaptive endpointing, it is reasonable to assume that features derived from the initial parts of an utterance can be fed to an online model.
We compare these features and how they impact the performance of our models.

\noindent\textbf{How much information do we need?}
Since an endpointing agent needs to choose a configuration before the entire input has been consumed, the task is to predict in advance whether an early cutoff is likely, rather than detecting an event that has already happened.
Clearly the more of the input the agent sees, the more accurate the predictions will be.
In simulations we investigate the effects of varying amounts of prior data for making a decision about the endpointing configuration.

\noindent\textbf{Can an online contextual bandit model be used in place of an offline-trained model?}
Finally, we note that a supervised learning framework is not suitable for online learning, and investigate whether an online model may be used instead.
Specifically, we adopt the contextual multi-armed bandit (CMAB) framework so that models can learn from reward signals based on latency and cutoff results, instead of ground-truth annotations.

In summary, we find that 
(1) target audio and partial ASR hypotheses based on the starts of utterances are most important;
(2) the more target audio data, the better the performance up to a point---with only about an initial 20\% of the data, an agent can reduce early cutoff without degrading latency---; 
and 
(3) online models such as deep CMAB are applicable to the endpointing task, reducing cutoffs while maintaining latency performance.

\section{Task \& Methodology}

\subsection{Task: Adaptive endpointing}
\label{subsec:p_ep}
Our task is to predict the optimal endpointing configuration for the speaker.
While there could be an unlimited number of configurations for each hyperparameter set, we limit our attention to this binary setting, where the focus is to reliably predict when the target utterance is better endpointed with the ``relaxed'' configuration, as opposed to the default or ``standard'' configuration.


\label{sec:task_methodology}

\subsection{Dataset}
\label{subsec:dataset}
For our study, we use de-identified data sampled from a voice-enabled assistant.
Using this collection of utterances, we then annotate each utterance with ground truth information using the following logic:
\begin{itemize}
    \item If an utterance is cut off early with the ``standard`` configuration, label the utterance as Class 1, meaning that the optimal configuration for the utterance is ``relaxed`` (positive class).
    \item Conversely, if an utterance is not cut off early, then the utterance is labeled ``standard``, or Class 0.
\end{itemize}
We split our collection of about 610 hours into training, development, and test splits with a ratio of 8:1:1, each with about 2.5\% positive labels (early cut-off with ``standard`` configuration). 
Overall, only about 0.02\% of the utterances are cut off early in both of the configurations.
The audio data is in English. 

\subsection{System Architecture \& Features}
\label{subsec:architecture_features}
\begin{figure}[t]
  \centering
  \includegraphics[width=\linewidth]{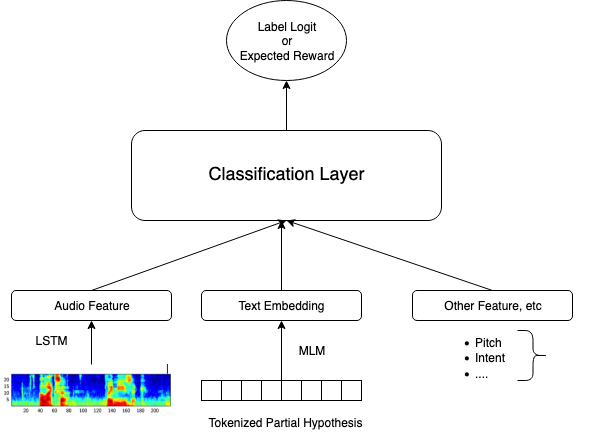}
  \caption{Shared model architecture for the supervised classifier and deep contextual bandit models. For the bandit models, the classification layer is replaced by a reward predictor for each action.}
  \label{fig:model}
\end{figure}

Our model architecture is an LSTM-based model proposed by Maas et al, which uses acoustic features and is pretrained to predict both end-of-utterance and voice activity \cite{Maas2018}.
We make necessary modifications by adding a last-frame pooling step, and add an MLM-based text encoder to embed partial text hypotheses.
The audio and hypothesis features are concatenated with additional (``Other'') features and fed to the classification layer for prediction.

\begin{table}[t]
  \caption{Input Features}
    \label{tab:features}
  \centering
  \begin{tabular}{ll}
    \hline
    \textbf{Name}      & \textbf{Description}                \\
    \hline\hline
    Audio                    &    extracted audio features                                    \\
    Hypothesis                   &     best-1 hypothesis from decoding                            \\
    Pause Duration                   &   time between wakeword and intent                               \\
    Wakeword Duration                   &  duration of wakeword                                \\
    Pitch Features                   &      paralinguistic features for intonation                            \\
    Intent Domain                   &      domain of the intentful utterance                            \\
    \hline
  \end{tabular}
\end{table}

The full set of features we experimented with is listed in Table~\ref{tab:features}.
Audio and hypothesis features represent the acoustic and semantic content of the utterance, respectively.
Language model features model the syntactic and semantic completeness of utterances and have been shown to boost the performance of endpointing models \cite{Ferrer2003}.
While transcription hypotheses are distinct from language model posteriors typically used in such studies, we take partial or complete transcriptions of the target audio as a proxy for language model predictions.
In addition, we also include some hand-crafted features that are relevant to endpointing based on prior work.
Wakeword and pause duration could be indicative of initial speaking rate or hesitation by the speaker, while past research shows that prosodic and paralinguistic features, such as pitch, are important for endpointing \cite{Ferrer2003,Maas2017, Ishimoto2017EndofUtterancePB, Liu2017TurnTakingEM, Thomas2012AcousticAD, Maier2017TowardsDE}.
Also, we assume oracle access to the intent domain, which refers to the category of the user's command. 
We included this feature based on our intuition that certain intents or commands are more likely to induce slower or more disfluent speech production, such as web search or question answering.

\subsection{Deep contextual multi-armed bandit (Deep CMAB) algorithm}
\label{subsec:dcmab}
A key disadvantage of the supervised approach is that it cannot be trained in an online manner, since training requires knowing the decoding result with the ``standard`` configuration, regardless of which decision it made during the prediction step, making it necessary to prepare the training set in advance.
On the other hand, online learning frameworks such as CMAB only require that the agent receive a reward signal, for the chosen action \cite{Agrawal2013ThompsonSF}.

Online bandit models such as Linear Thompson sampling have been successfully applied in online prediction settings such as recommendation systems \cite{Moerchen2020}, and recent studies show that moving from offline models trained on custom datasets to online models trained using implicit signal can provide significant improvement in performance and cost \cite{Yu2021}.

Thus, we adopt the CMAB approach for our online agents.
Specifically, we adopt the recent deep implementation of the CMAB, instead of popular linear algorithms such as Linear Thompson sampling \cite{Russo2018}.
We find that deep bandits are better suited to our task both for their representational power and for handling audio and their batched training capability.
While there are different algorithms for deep bandits, such as proposed in \cite{Zhang2021neural, Riquelme2018}, we adopt the relatively simple framework proposed in \cite{Collier2018}.
One difference between \cite{Collier2018} and our approach is that instead of updating the model periodically, we update our model after seeing each batch of examples.
Specifically, we adopt concrete dropout in place of conventional dropout weights for neural network model training.
Gal et al.\ show that concrete dropout allows the model to calibrate the amount of exploration naturally, as training progresses \cite{Gal2017}.

\SetKwComment{Comment}{/* }{ */}
\begin{algorithm}[t]
\KwData{Set of utterances $S$, Neural Network $f$}
\For{$\texttt{utterances} \in S$}{
    \For{$a_i \in \texttt{Actions}$}{
        $\texttt{rewards}_i \gets f(s, a_i))$
    }
    $\texttt{chosen} \gets \textbf{\texttt{choose\_action}}\textbf{(rewards)}$\\
    $\texttt{real\_reward} \gets \texttt{decode}\texttt{(s, chosen)}$\\
    $f \gets \textbf{update}\texttt{($f$, real\_reward, rewards)}$\\
    \If{terminate}{break}
}
\caption{Deep Contextual Bandit Pseudocode}\label{alg:deep_bandit}
\end{algorithm}

\begin{table*}[t]
  \caption{Endpointing metrics obtained with different features in an idealized setting where the features are computed from the whole target audio. Relative results (indicated by $\pm$) use the Standard Only as a baseline.
  TM95 refers to a trimmed mean (lower 95th percentile). DTM95:99 refers to a doubly-trimmed mean (95th to 99th percentile).} 
  \label{tab:compare_features_idealized}
\resizebox{\textwidth}{!}{%
\begin{tabular}{lccc||cccccc}
\hline
Metrics / Model & Standard Only & Relaxed Only & Oracle Result & \begin{tabular}[c]{@{}c@{}}Target\\ Audio \end{tabular}& \begin{tabular}[c]{@{}c@{}}Target\\ Hypothesis\end{tabular} & \begin{tabular}[c]{@{}c@{}}Intent\\ Domain \end{tabular} & \begin{tabular}[c]{@{}c@{}}Wakeword\\ Duration\end{tabular} & \begin{tabular}[c]{@{}c@{}}Pitch\\ Features\end{tabular} & \begin{tabular}[c]{@{}c@{}}Pause\\ duration\end{tabular} \\ \hline\hline
Accuracy (\%) & 97.46 & 2.75 & 100 & 87.48 & 97.68 & 21.11 & 95.23 & 94.40 & 67.43 \\
Precision (\%) & NA & 2.75 & 100 & 17.19 & 53.50 & 2.29 & 0.40 & 3.66 & 2.46 \\
Recall (\%) & NA & 100 & 100 & 91.57 & 84.83 & 68.41 & 0.29 & 4.07 & 28.67 \\
F1 score & NA & 5.36 & 100 & 28.95 & 65.61 & 4.44 & 0.34 & 3.86 & 4.53 \\
Early EP rate & - & 0.07 & 0.05 & 0.29  & 0.43 & 0.89  & 2.77  & 2.65  & 1.94  \\
Early EP rate change & - & -97.24\% & -98.03\% & -88.58\% & -83.07\% & -64.96\% & +9.06\% & +4.33\% & -23.62\% \\
Latency (TM95) & - & +323.42\% & +1.65\% & +29.75\% & +3.03\% & +257.58\% & +0.28\% & +3.03\% & +95.59\% \\
Latency (DTM95:99) & - & +108.92\% & +46.51\% & +95.66\% & +67.47\% & +106.99\% & +27.59\% & +64.34\% & +100.24\%
\\ \hline
\end{tabular}%
}

\end{table*}

The pseudocode for our algorithm is given below as Algorithm~\ref{alg:deep_bandit}. The neural network model uses the architecture shown in Figure~\ref{fig:model}, while for \textbf{\texttt{choose\_action}}, we take the greedy action with argmax, and \textbf{\texttt{update}} the model through stochastic gradient reward.
In the implementation, our model predicts rewards for both actions simultaneously, rather than predicting a reward given an action.
For the bandit model, the reward signal is computed as a linear combination of latency (in ms) and cutoff (indicator variable), which is then used in the reward prediction loss calculation (mean squared error loss).
The mixing weights are hyperparameters that we find using experiments on held-out development data.
Intuitively, the bandit model tries to predict the expected reward of each configuration, and chooses a configuration based on the reward, while the supervised classifier directly outputs the predicted optimal configuration.





\section{Experiments}
\label{sec:experiments}

\subsection{Experiment setup}
\label{subsec:ex_setup}
Beyond accuracy measures, we also evaluate the performance of our models using the following metrics:
(1) Early endpointing rate (Early EP rate), which measures the fraction of times the endpointer triggers before the end of the last utterance is reached.
(2) Trimmed mean 95 (TM95) of latency (ms), average of the lower 95th percentile of the data
(3) Double-trimmed mean 95 (DTM95:99) of latency (ms), average of the interval (inclusive) between the 95th and 99th percentiles of the data.

To contextualize and better compare the performance of our models, we also measure the results of several baselines.
\textbf{Standard Only}, the default baseline model, always chooses the ``standard'' configuration, while \textbf{Relaxed Only} always chooses the ``relaxed'' configuration. 
Lastly, we also consider the \textbf{Oracle Model}, which always outputs the optimal configuration choice, giving an upper bound of achievable performance.


\subsection{Which features are most useful?}
\label{subsec:result1}

To study which features are important for adaptive endpointing, we conduct utterance-wise endpointing experiments in an idealized setting.
That is, we assumed that the adaptive endpointing model will have the full length of the target audio and the features derived from the full audio as inputs.
In Table~\ref{tab:compare_features_idealized}, we note that target audio and hypothesis features achieve largest cutoff reduction, followed by intent domain, which incurs significant latency degradation.

We note that the top-performing features (target audio, target hypothesis, and intent domain) require processing of the target audio utterance. 
On the other hand, features that can be obtained without the intent-carrying portion of an utterance (wakeword duration, pitch features) perform poorly, showing that it is difficult to reliably predict the overall utterance pattern just from the paralinguistic features derived from the initial parts of the utterance. 

\textbf{Conclusion}: Hypothesis and audio features are most informative. In an idealized setting where the whole target utterance is used for prediction, hypothesis and audio features lead to 80\% and 20\% relative cutoff rate reduction, respectively, with no TM95 degradation and Oracle-level DTM95:99 latency. 

\vspace{-0.25cm}

\begin{figure}[t]
\centering
\begin{tabular}{ll}
\includegraphics[width=0.45\textwidth]{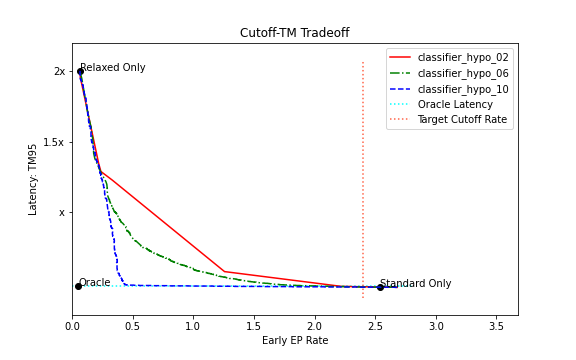}
\\
\includegraphics[width=0.45\textwidth]{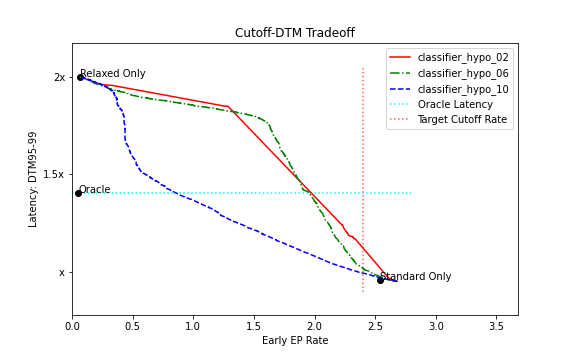}
\end{tabular}
\caption{Comparing hypothesis features with varying portions on cutoff-latency tradeoff curves.}
\label{fig:compare_features_hypothesis}
\end{figure}

\subsection{How much information do we need?}
\label{subsec:result2}

In practice, consuming the whole target audio is unrealistic.
Thus, we relax this assumption by assuming our models now see data corresponding to the first $X$\% of the utterance.
To derive the input features for this setup, we first process the full target utterance to derive corresponding audio and hypothesis features. 
Then, we take the first $X$\% of the resulting features and feed them to our model.
We conducted experiments to compare bandit models trained with all types of features, but we chose to include only the two features with the best performance, for brevity and legibility.

Figure~\ref{fig:compare_features_hypothesis} shows the latency vs.\ early EP trade-off curves plotted for supervised classifiers with target hypothesis features with varying amount (20\%, 60\%, 100\%) of tokens.
First, we observe that the cutoff vs.\ DTM95:99 curve has a ``worse`` trade-off curve since more latency degradation is required to achieve the same amount of early cutoff reduction. 
This is because by definition of DTM95:99 is more sensitive to changes in the tail of the latency distribution.
However, we note that even the 20\% model achieves no TM95 latency degradation and only $\sim$20\% DTM95:99 latency degradation, as indicated by the intersection between the blue curves and the dotted red line (target cutoff rate).

Moreover, we confirm that using a larger portion of the target features improves the performance.
In Figure~\ref{fig:compare_features_hypothesis}, the 100\% model achieves significantly better trade-offs for both TM95 and DTM95:99.
This is a confirmation of the intuition that the latter parts of the audio provide more information about which endpointing configuration is optimal for the audio.

\textbf{Conclusion:} By using only the initial 20\% of the data, we can reduce Early-EP rate by 5\%, with no degradation in TM95 and 20\% relative DTM95:99 degradation over the baseline.
However, we also find that the latter parts of the contain valuable information about optimal endpointing configurations.
\vspace{-0.25cm}

\subsection{Can an online contextual bandit model be used in place of an offline-trained model?}
\label{subsec:result_3}


\begin{figure}[!h]
\centering
\begin{tabular}{ll}
\includegraphics[width=0.45\textwidth]{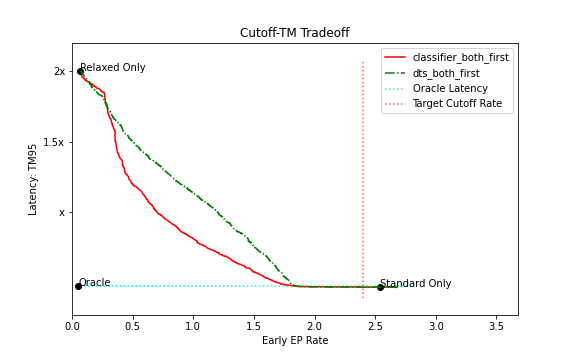}
\\
\includegraphics[width=0.45\textwidth]{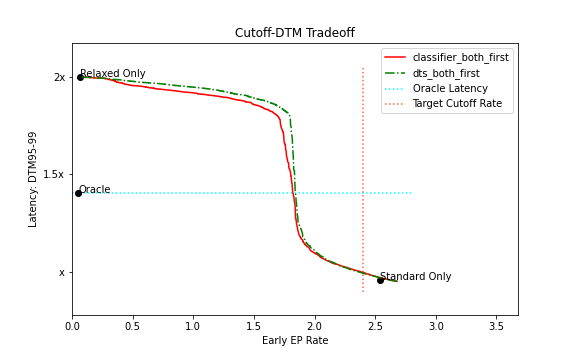}
\vspace{-0.25cm}
\end{tabular}

\caption{Comparing supervised and bandit models on first segment input features.}
\label{fig:compare_models_first}
\end{figure}

Since the offline setup required by the supervised model is not ideal, as discussed in Section~\ref{subsec:dcmab}, we investigate if a deep CMAB model can also meet the objective of reducing early cutoff without degrading latency, by comparing the supervised and bandit models.
Furthermore, we compare these predictive models in a more realistic setting where the total length of the target audio is not known beforehand.
(In previous experiments, we either assumed an idealized setting, with access to the full target utterance, or knowing the utterance length in advance.)
Hence, we extracted the initial segment of an utterance containing the wake word, and use the audio and hypothesis corresponding only to that segment. 
%
On average, the time fraction of that initial segment is $\sim$ 30\% of the full utterance.
We consider this setup a simulation of when a speculative listener is first activated, and retrieves a partial utterance and decoding result (hypothesis) to the endpointing model.

Figure~\ref{fig:compare_models_first} shows the trade-off curves for the supervised and bandit models.
While the bandit model (dts\_both\_first) achieves slightly worse trade-offs for both TM95 and DTM95:99, both models achieve target cutoff reductions with little (DTM95:99) to no (TM95) sacrifice for latency measures.
We note that both supervised and bandit models achieve a significant (2.5 $\rightarrow$ 1.8, $\sim$30\%) Early-EP rate reduction without any TM95 latency degradation, while a small degradation in DM95:99 latency is observed.

\textbf{Conclusion:} Deep contextual bandits can reduce cutoff rate by 5\% without TM95 latency degradation, and $\sim$20\% DTM95:99 degradation. The more target data is available, the more the gap between supervised classifier and bandit model narrows.

\vspace{-0.25cm}
\section{Conclusions}
\label{sec:conclusion}
We have proposed adaptive endpointing as a framework for dynamically choosing an optimal endpointing configuration, based on features derived from each input utterance.
By implementing a static supervised classifier for endpointing configuration, we show that utterance-level selection of the locally best endpointing configuration leads to reduction in early cutoff rate, while keeping latency degradation small.
We also show that an online model can be trained without having access to ground truth data. 
For this purpose, a deep contextual multi-armed bandit (CMAB) model combines the efficiency of Bayesian exploration with the representational power of neural networks,
does not require ground truth annotation, and can be adapted to utilize a variety of reward signals that may be available in an online deployment setting.
We find that audio and text features derived from the target utterance are most important for endpointing, and online-trained deep CMAB models can be used in place of impractical offline supervised classifiers, while still reducing early cutoff without latency degradation.



\bibliographystyle{IEEEbib}
\bibliography{strings,refs}

\begin{thebibliography}{10}

\bibitem{Ferrer2003}
L.~Ferrer, E.~Shriberg, and A.~Stolcke,
\newblock ``A prosody-based approach to end-of-utterance detection that does
  not require speech recognition,''
\newblock in {\em Proc.\ IEEE ICASSP}, 2003, vol.~1, pp. I--I.

\bibitem{Arsikere2014ComputationallyefficientEF}
Harish Arsikere, Elizabeth Shriberg, and Umut Ozertem,
\newblock ``Computationally-efficient endpointing features for natural spoken
  interaction with personal-assistant systems,''
\newblock in {\em Proc.\ IEEE ICASSP}, 2014, pp. 3241--3245.

\bibitem{Li2022}
Siyan Li, Ashwin Paranjape, and Christopher~D. Manning,
\newblock ``When can {I} speak? {P}redicting initiation points for spoken
  dialogue agents,'' arXiv:2208.03812, 2022.

\bibitem{Chang2022}
Shuo-yiin Chang, Bo~Li, Tara~N. Sainath, Chao Zhang, Trevor Strohman, Qiao
  Liang, and Yanzhang He,
\newblock ``Turn-taking prediction for natural conversational speech,'' 2022.

\bibitem{Schlangen2006FromRT}
David Schlangen,
\newblock ``From reaction to prediction: Experiments with computational models
  of turn-taking,''
\newblock in {\em Proc.\ Interspeech}, Pittsburgh, 2006, pp. 2010--2013.

\bibitem{Zhao2021}
Yingzhu Zhao, Chongjia Ni, Cheung-Chi Leung, Shafiq Joty, Eng~Siong Chng, and
  Bin Ma,
\newblock ``Preventing early endpointing for online automatic speech
  recognition,''
\newblock in {\em Proc.\ IEEE ICASSP}, 2021, pp. 6813--6817.

\bibitem{Lu2022}
Liang Lu, Jinyu Li, and Yifan Gong,
\newblock ``Endpoint detection for streaming end-to-end multi-talker {ASR},''
\newblock in {\em Proc.\ IEEE ICASSP}, 2022, pp. 7312--7316.

\bibitem{Huang2022E2ESJ}
W.~Ronny Huang, Shuo yiin Chang, David Rybach, Rohit Prabhavalkar, Tara~N.
  Sainath, Cyril Allauzen, Cal Peyser, and Zhiyun Lu,
\newblock ``{E2E Segmenter}: Joint segmenting and decoding for long-form
  {ASR},''
\newblock in {\em Proc.\ Interspeech}, 2022.

\bibitem{Maas2018}
Roland Maas, Ariya Rastrow, Chengyuan Ma, Guitang Lan, Kyle Goehner, Gautam
  Tiwari, Shaun Joseph, and Björn Hoffmeister,
\newblock ``Combining acoustic embeddings and decoding features for
  end-of-utterance detection in real-time far-field speech recognition
  systems,''
\newblock in {\em Proc.\ IEEE ICASSP}, 2018, pp. 5544--5548.

\bibitem{Jayasimha2021PersonalizingSS}
Aditya Jayasimha and Periyasamy Paramasivam,
\newblock ``Personalizing speech start point and end point detection in {ASR}
  systems from speaker embeddings,''
\newblock {\em Proc.\ IEEE Spoken Language Technology Workshop}, pp. 771--777,
  2021.

\bibitem{Ding2022PersonalV2}
Shaojin Ding, Rajeev~Vijay Rikhye, Qiao Liang, Yanzhang He, Quan Wang, Arun
  Narayanan, Tom O'Malley, and Ian McGraw,
\newblock ``Personal {VAD} 2.0: Optimizing personal voice activity detection
  for on-device speech recognition,''
\newblock in {\em Proc.\ Interspeech}, 2022.

\bibitem{Munkhdalai2022}
Tsendsuren Munkhdalai, Khe~Chai Sim, Angad Chandorkar, Fan Gao, Mason Chua,
  Trevor Strohman, and Françoise Beaufays,
\newblock ``Fast contextual adaptation with neural associative memory for
  on-device personalized speech recognition,''
\newblock in {\em Proc.\ IEEE ICASSP}, 2022, pp. 6632--6636.

\bibitem{Sathyendra2022}
Kanthashree~Mysore Sathyendra, Thejaswi Muniyappa, Feng-Ju Chang, Jing Liu,
  Jinru Su, Grant Strimel, Athanasios Mouchtaris, and Siegfried Kunzmann,
\newblock ``Contextual adapters for personalized speech recognition in neural
  transducers,''
\newblock in {\em Proc.\ IEEE ICASSP}, 2022.

\bibitem{Maas2017}
Roland Maas, Ariya Rastrow, Kyle Goehner, Gautam Tiwari, Shaun Joseph, and
  Bjorn Hoffmeister,
\newblock ``Domain-specific utterance end-point detection for speech
  recognition,''
\newblock in {\em Proc.\ Interspeech}, 08 2017, pp. 1943--1947.

\bibitem{Ishimoto2017EndofUtterancePB}
Yuichi Ishimoto, Takehiro Teraoka, and Mika Enomoto,
\newblock ``End-of-utterance prediction by prosodic features and
  phrase-dependency structure in spontaneous {Japanese} speech,''
\newblock in {\em Proc.\ Interspeech}, 2017.

\bibitem{Liu2017TurnTakingEM}
Chaoran Liu, Carlos~Toshinori Ishi, and Hiroshi Ishiguro,
\newblock ``Turn-taking estimation model based on joint embedding of lexical
  and prosodic contents,''
\newblock in {\em Proc.\ Interspeech}, 2017.

\bibitem{Thomas2012AcousticAD}
Samuel Thomas, Sri Harish~Reddy Mallidi, Thomas Janu, Hynek Hermansky, Nima
  Mesgarani, Xinhui Zhou, Shihab~A. Shamma, Tim Ng, Bing Zhang, Long Nguyen,
  and Spyridon Matsoukas,
\newblock ``Acoustic and data-driven features for robust speech activity
  detection,''
\newblock in {\em Proc.\ Interspeech}, 2012.

\bibitem{Maier2017TowardsDE}
Angelika Maier, J.~Hough, and David Schlangen,
\newblock ``Towards deep end-of-turn prediction for situated spoken dialogue
  systems,''
\newblock in {\em Proc.\ Interspeech}, 2017.

\bibitem{Agrawal2013ThompsonSF}
Shipra Agrawal and Navin Goyal,
\newblock ``Thompson sampling for contextual bandits with linear payoffs,''
\newblock in {\em Proc.\ ICML}, 2013.

\bibitem{Moerchen2020}
Fabian Moerchen, Patrick Ernst, and Giovanni Zappella,
\newblock ``Personalizing natural language understanding using multi-armed
  bandits and implicit feedback,''
\newblock in {\em Proc.\ 29th ACM International Conference on Information and
  Knowledge Management}, New York, NY, USA, 2020, CIKM '20, p. 2661–2668.

\bibitem{Yu2021}
Ge~Yu, Chengwei Su, and Emre Barut,
\newblock ``Introducing deep reinforcement learning to {NLU} ranking tasks,''
\newblock in {\em Proc.\ IEEE ICASSP}, 2021.

\bibitem{Russo2018}
Daniel Russo, Benjamin Roy, Abbas Kazerouni, Ian Osband, and Zheng Wen,
\newblock ``A tutorial on {Thompson} sampling,''
\newblock {\em Foundations and Trends in Machine Learning}, vol. 11, no. 1, pp.
  1--96, 2018.

\bibitem{Zhang2021neural}
Weitong Zhang, Dongruo Zhou, Lihong Li, and Quanquan Gu,
\newblock ``Neural {Thompson} sampling,''
\newblock in {\em Proc.\ International Conference on Learning Representations},
  2021.

\bibitem{Riquelme2018}
Carlos Riquelme, George Tucker, and Jasper Snoek,
\newblock ``Deep {Bayesian} bandits showdown: An empirical comparison of
  bayesian deep networks for {Thompson} sampling,''
\newblock in {\em Proc.\ 6th International Conference on Learning
  Representations, {ICLR} 2018, Vancouver, BC, Canada, April 30 - May 3, 2018,
  Conference Track Proceedings}. 2018, OpenReview.net.

\bibitem{Collier2018}
Mark Collier and Hector~Urdiales Llorens,
\newblock ``Deep contextual multi-armed bandits,'' arXiv:1807.09809, 2018.

\bibitem{Gal2017}
Yarin Gal, Jiri Hron, and Alex Kendall,
\newblock ``Concrete dropout,''
\newblock in {\em Proc.\ 31st International Conference on Neural Information
  Processing Systems}, Red Hook, NY, USA, 2017, NIPS'17, p. 3584–3593, Curran
  Associates Inc.

\end{thebibliography}

\end{document}